\documentclass[a4paper,fleqn,usenatbib]{mnras}
\twocolumn
\usepackage{mathptmx}

\usepackage[T1]{fontenc}
\usepackage{ae,aecompl}

\usepackage{graphicx}	
\usepackage{amsmath}	
\usepackage{amssymb}	

\title[Formation of the absorption line 21-cm in Dark Ages]{Formation of the hydrogen line 21-cm in Dark Ages and Cosmic Dawn: dependences on cosmology and first light}

\author[B. Novosyadlyj et al.]{
Bohdan Novosyadlyj$^{1,2}$,\thanks{E-mail: bnovos@gmail.com}
Yurij Kulinich$^2$, Gennadi Milinevsky$^{1,3}$, Valerii Shulga$^{1,4}$
\\ \\
$^1$Jilin University, Qianjin Street 2699, Changchun, 130012, P.R.China,\\
$^2$Ivan Franko National University of Lviv, Kyryla i Methodia str., 8, Lviv, 79005, Ukraine \\
$^3$Taras Shevchenko National University of Kyiv, 64/13, Volodymyrska str.,  01601 Kyiv, Ukraine \\
$^4$Institute of Radio Astronomy of NASU, 4 Mystetstv str., 61002 Kharkiv, Ukraine}
\date{Accepted XXX. Received YYY; in original form ZZZ}
\pubyear{2016}

\begin{document}
\label{firstpage}
\pagerange{\pageref{firstpage}--\pageref{lastpage}}
\maketitle

\begin{abstract}
We analyze the formation of the redshifted hyperfine structure line 21-cm of hydrogen atom in the Dark Ages, Cosmic Dawn, and Reionization epochs. The evolution of the global differential brightness temperature in this line was computed  to study its dependence on the values of cosmological parameters and physical conditions in the intergalactic medium. Variations of the depth of the Dark Ages absorption line at $z\sim80$ with variations of the cosmological parameters $\Omega_b$, $\Omega_{cdm}$, $\Omega_{\Lambda}$, $\Omega_K$ and $H_0$ are studied.  The standard model with post-Planck parameters predicts a value of the differential brightness temperature in the center of the absorption line $\sim$30-50 mK. The profile of this line can be quite another in the non-standard cosmological models, which include the annihilating or decaying dark matter, a primordial stochastic magnetic field, etc. It can be shallower or be an emission bump instead of an absorption trough. It is also shown that the position and depth of the Cosmic Dawn absorption line formed at 10<z<30, due to the Wouthuysen-Field effect, is mainly defined by the spectral energy distribution of the first sources of light. If reionization occurs at $z_{ri}=7\pm1$, then the differential brightness temperature in the center of this line is $\sim$80 mK. During the reionization, the emission with an amplitude of $\sim$20 mK is possible. It is also shown that the temperature, density, and degree of ionization of the baryonic component are decisive in calculating the intensity of the 21-cm absorption/emission line from these epochs.
\end{abstract}

\begin{keywords}
cosmology: theory -- large-scale structure of Universe -- dark energy
\end{keywords}


\section{Introduction}
Recent data on massive galaxies and quasars at high redshifts have heightened interest in the early epochs when the first luminous objects in our Universe began to form. An important information channel about the state of baryonic matter in this period is the redshifted line 21-cm of neutral hydrogen (see reviews  \cite{Barkana2001,Fan2006,Furlanetto2006,Bromm2011,Pritchard2012}). The earliest signal from forming halos in Dark Ages can be received in this spectral line as well \citep{Iliev2002,Iliev2003,Furlanetto2006a,Shapiro2006,Kuhlen2006,Novosyadlyj2020}. The physical conditions of hydrogen gas, the excitation and ionization states during the Dark Ages and Cosmic Dawn epochs in the standard cosmological models and scenarios of the first light sources formation are well studied. The known spectral features include the absorption wide lines redshifted to $\sim20$ MHz at $z\sim80$ and  $\sim70-130$ MHz at $z\sim20-10$, and the emission line before complete reionization. The second absorption line is caused by the Wouthuysen-Field effect \citep{Wouthuysen1952,Field1958,Field1959} and is determined by the spectral energy distribution (SED) of the first sources of light (the first light). Non-contradictory models of the first light sources in the standard cosmological model predict a line depth that does not exceed $\sim 250$ mK of brightness temperature \citep{Cohen2017}. The first and to date only detection of this line in the Experiment to Detect the Global Epoch of Reionization Signature  experiment (EDGES) \citep{Bowman2018} indicates an unusually shaped profile and an unexpectedly large depth centred on 78 MHz, which is $\sim$3-4 times deeper than expected in the standard cosmology. The explanations go out of the bounds of standard cosmology, including additional mechanisms for cooling the baryons, excess radio background at high redshifts, viscous dark energy, and so on \citep{Barkana2018,Ewall-Wice2018,Halder2022}. Another explanation consists of the challenge of measuring useful signals at the huge foreground and decreasing the systematic errors \citep{Hills2018}. The recent non-detection of a signal from the Cosmic Dawn epoch in the Shaped Antenna measurement of the background RAdio Spectrum 3 (SARAS3) experiment \cite{Singh2022} supports the last assumption: its spectrum has not the feature found by \citep{Bowman2018} and rejects their best-fitting profile with 95.3\% confidence. But SARAS3 \citep{Singh2022} nothing sad about the spectral feature of the Cosmic Dawn signal in the range of 55–85 MHz, therefore, the predictions of signal in this range by the standard model are actual, and their measurements are even more urgent. 

The most prominent spectral feature of the redshifted 21-cm line is the absorption line formed in Cosmic Dawn when the scattering of $Ly\alpha$-radiation of the first sources of light affects the populations of hyperfine structure levels of ground state hydrogen \citep{Field1959,Hirata2006}. Theoretical aspects of its formation are comprehensively studied since it can bring us information about the first sources of light, the first stars, galaxies and phenomena related to their origin. The traditional approaches to the modeling of the $Ly\alpha$-coupling are based on the phenomenological connections between stars and galaxies formation rates and intensity of $Ly\alpha$-radiation (see, for example, review \cite{Furlanetto2006}. In this paper we analyse the dependence of spectral features of the 21-cm line of Dark Ages, Cosmic Dawn, and Reionization epochs on cosmological parameters and models of the first light in the non-traditional approach: the intensity of $Ly\alpha$-radiation we deduce for given SED of the first light using the observational constraints on the Reionization epoch. By the variations time of appearance of the first light, rate of increasing its intensity  and ratio of $Ly\alpha$-photons to ionization ones we estimate the possible spectral position and intensity of the 21-cm line taking into account the observational constraints on $x_{HII}(z)$ at $6\le z\le20$ \citep{Planck2020a,Planck2020b,Bouwens2015,Banados2018,Davies2018,Mason2018}.

The outline of the paper is as follows. In Section 2 we describe the models of the energy distribution of the light from the first sources that appeared in the Cosmic Dawn, and state of atomic hydrogen from cosmological recombination to reionization. In Section 3 we analyse the dependences of the position and depth of the Dark Ages absorption line to cosmological parameters and additional heating and cooling of the baryonic component. In Section 4 we estimate the spectral and redshift position and intensity of the 21-cm line depending on the SED of the first light and its evolution. The results are summarised in Section 5.

\section{State of atomic hydrogen from Cosmological Recombination to Reionization}

\begin{figure}
\includegraphics[width=0.5\textwidth]{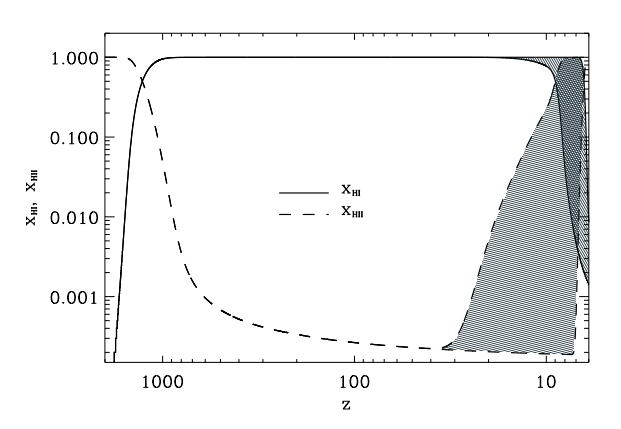}
\caption{Fractions of neutral and ionized hydrogen from the cosmological recombination  at z=2000 up to complete reionization at z=6. Shaded range is 2$\sigma$ confidence range following from the Plank low-l polarization data \citep{Planck2020a,Planck2020b}.}
\label{frac}
\end{figure}

\begin{figure}
\includegraphics[width=0.5\textwidth]{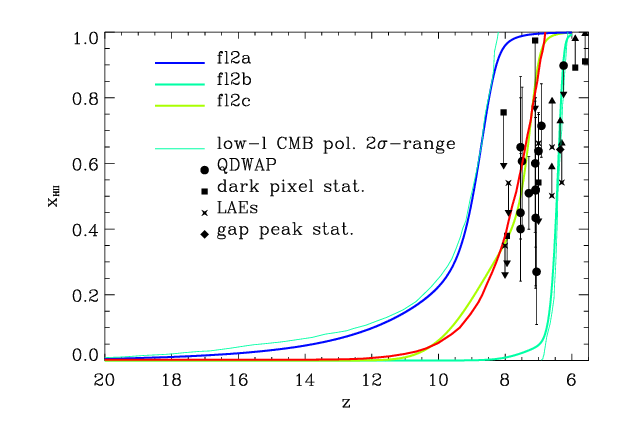}
\caption{Fraction of ionized hydrogen at Cosmic Dawm and Reionization epochs from cosmological and astrophysical observational data.}
\label{rei}
\end{figure}

The neutral hydrogen atoms were the dominant component after the Cosmological Recombination epoch and before Reionization one. During the Dark Ages epoch, it is about 99.98\%, decreases when the UV radiation of the first sources of light appear in the Cosmic Dawn epoch and sharply decreases during Reionization one\footnote{Molecular hydrogen fractions are a few orders lower (\cite{Novosyadlyj2022} and citations therein) and can be omitted here from consideration.}. The cosmological recombination is well studied both theoretically (\cite{Seager1999,Seager2000} and citing therein) and instrumentally in ground-based, stratospheric and space observations of the cosmic microwave background radiation. We know about Dark Ages and Cosmic Dawn mainly from the theory and have a few scenarios of forming the first light sources without direct observational support. The study of reionization has a long history (see overviews \cite{Choudhury2022,Gnedin2022} and citing therein). According to the currently popular models, the UV radiation of the first stars of the early galaxies reionized hydrogen atoms progressively throughout the entire Universe at redshift $6<z<12$, while helium atoms have been reionized by hard UV and X- radiation of quasars at $2<z<6$ \citep{Planck2020a,Gnedin2022}. Figure \ref{frac} illustrates the evolution of $x_{HI}\equiv n_{HI}/n_H$ and $x_{HII}\equiv n_{HII}/n_H$, where $n_H\equiv n_{HI}+n_{HII}$, from the beginning of hydrogen recombination at $z=2000$ up to complete reionization at $z=6$. The inaccuracy of the calculation of cosmological recombination using the code RecFast \cite{Seager1999,Seager2000} in the framework of the given cosmological model is not larger than 1-3\%, and the current uncertainties of reionization epoch are shown by the shaded zone, which is 2$\sigma$ confidence range following from the Plank low-l polarization measurements \citep{Planck2020a,Planck2020b}. Such measurements are very demanding since the amplitude of the E-mode polarization power spectrum at low multipoles is lower by more than two orders of magnitude than the amplitude of the temperature anisotropy power spectrum. Other probes of the reionization epoch based on the spectral features of the most distant quasars and galaxies give $x_{HI}$-value in the shaded range. They are presented in Fig. \ref{rei}, where Planck 2$\sigma$-range of constraints on $x_{HII}$ \citep{Planck2020a} is shown by the thin green solid lines, and the median value $\overline{x}_{HII}(z)$ \citep{Glazer2018} by the thick red solid line. The astrophysical observational data are shown also. There are data on $x_{HII}(z)=1-x_{HI}(z)$ derived from the dark pixel statistics (squares, \cite{McGreer2015}), the gap/peak statistics (diamonds, \cite{Gallerani2008}), the damping wing absorption profiles in the spectra of quasars (QDWAP) (circles, \cite{Schroeder2013, Greig2017,Mortlock2011,Davies2018,Banados2018,Greig2022}), the redshift-dependent prevalence of $Ly\alpha$ emitters (LAEs) (4-fold stars, \cite{Schenker2014,Mason2018,Mason2019,Ouchi2010}).
 
\begin{table}
\begin{center}
\caption{Parameters of models of the first ligh 1.}  
\begin{tabular} {c|ccccc}
\hline
\hline
   \noalign{\smallskip}
Model&$T_*$ (K)&$\alpha_{fl}$&$z_{fl}$&$a_{fl}$&$b_{fl}$ \\
 \noalign{\smallskip} 
\hline
   \noalign{\smallskip} 
 fl1a&5000&$5.0\cdot10^{-8}$&0.2&5.0&0.7\\
    \noalign{\smallskip}    
  \hline 
 fl1b&10000&$6.0\cdot10^{-15}$&2.4&4.0&1.4\\
    \noalign{\smallskip}    
  \hline
 fl1c&20000&$1.0\cdot10^{-19}$&2.5&3.15&1.4\\
    \noalign{\smallskip}    
  \hline  
  \hline
\end{tabular}
\end{center}
\label{pmfl1}
\end{table}
\begin{figure*}
\includegraphics[width=0.32\textwidth]{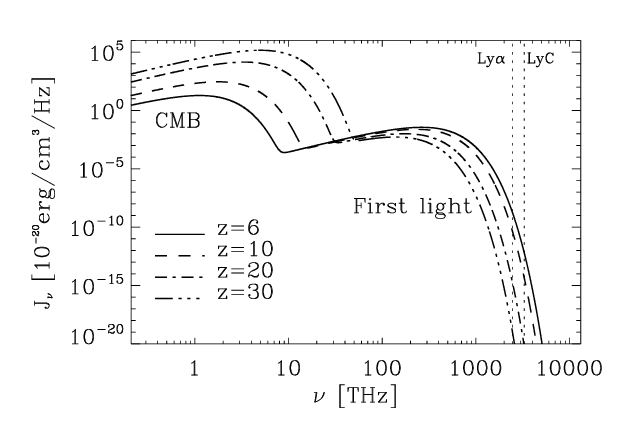}
\includegraphics[width=0.32\textwidth]{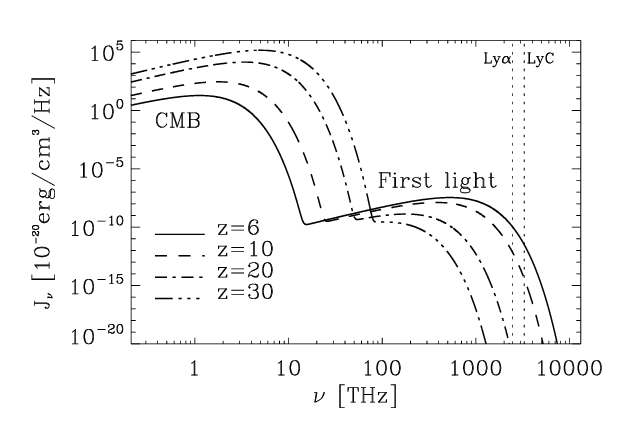}
\includegraphics[width=0.32\textwidth]{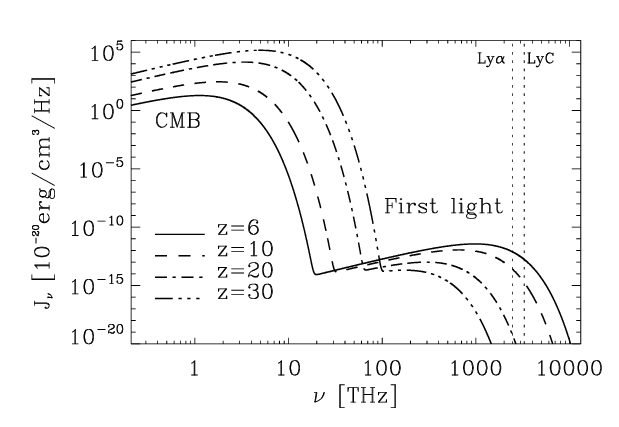}
\caption{The evolution of SEDs in the models fl1a, fl1b and fl1c (from left to right) for the Cosmic Dawn and Reionization epochs with parameters presented in Tab. 1.}
\label{sedfl1}
\end{figure*}

\begin{figure}
\includegraphics[width=0.5\textwidth]{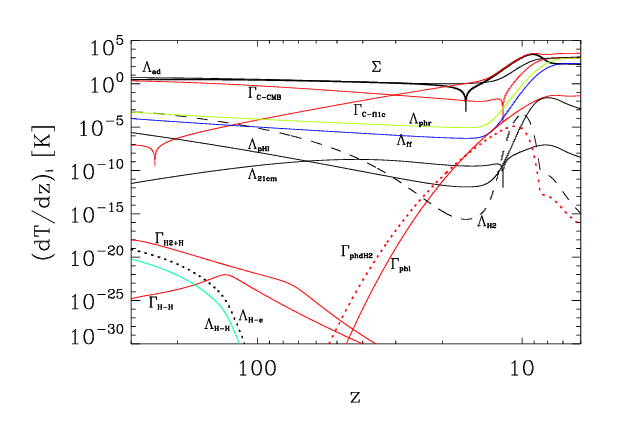}
\includegraphics[width=0.5\textwidth]{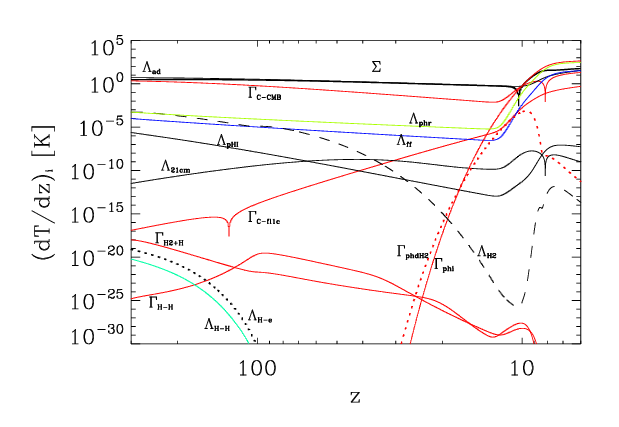}
\includegraphics[width=0.5\textwidth]{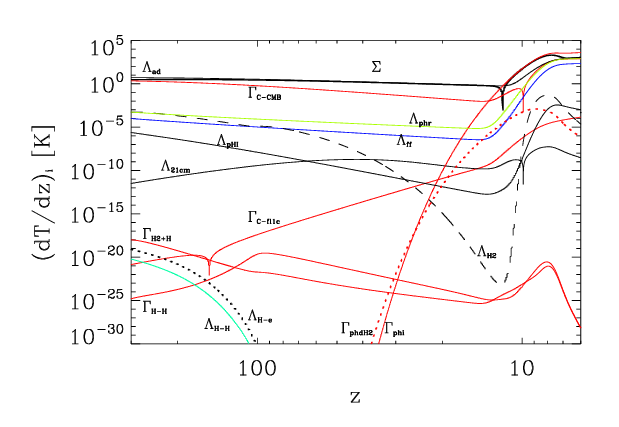}
\caption{Heating (red lines) and cooling functions in units of Kelvins ($\left(dT/dz\right)_i\equiv2\Gamma_i/\left(3n_{tot}k_B(1+z)H(z)\right)$, $\left(dT/dz\right)_i\equiv2\Lambda_i/\left(3n_{tot}k_B(1+z)H(z)\right)$) for models of the first light with parameters presented in Tab. 1.}
\label{hc}
\end{figure}

\begin{figure}
\includegraphics[width=0.5\textwidth]{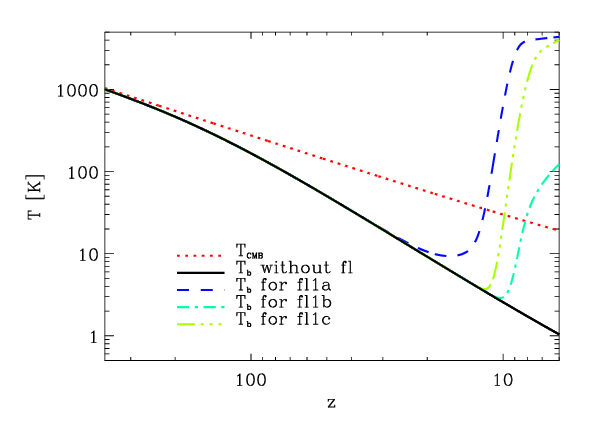}
\caption{The evolution of the temperature of baryonic matter $T_b$ from the Dark Ages epoch, through the Cosmic Dawn one up to complete hydrogen reionization for the first light models with parameters presented in Tab. 1.}
\label{figTb}
\end{figure}

In this paper, we analyse a few models of the first light with different SEDs, which provide the evolution of $x_{HII}(z)$ shown in Fig. \ref{frac}-\ref{rei} and their influence on the absorption/emission line 21-cm of neutral hydrogen before and during reionization. To analyze the allowable level of illumination in the inter-proto-galaxy medium of the Cosmic Dawn epoch we assume that the sources of the first light are thermal. We consider here the thermal models of the first light described by the Planck function with temperature $T_{fl}$, which is a smooth function of redshift, and dilution coefficient $\alpha_{fl}$. The spectral energy density of total radiation at some equidistance from them is as follows
\begin{eqnarray}
J_{fl}(\nu)&=&\frac{4\pi}{c}\left[B(\nu;T_{CMB})+\sum_i\alpha^{(i)}_{fl}B(\nu;T^{(i)}_{fl})\right], \label{jnu} \\
T_{fl}^{(i)}&=&T_*^{(i)}\tanh{\left[a_{fl}^{(i)}\left(\frac{1+z_{fl}^{(i)}}{1+z}\right)^{b_{fl}^{(i)}}\right]},\label{tfl}
\end{eqnarray} 
where $B(\nu,T)$ is the Planck function. The coefficients $\alpha_{fl}$, $a_{fl}$, $b_{fl}$ and $z_{fl}$ are fitting ones to obtain the $x_{HII}(z)$ matching the observational data for given $T_{*}$. 

We consider three specific reionization z-tracks: early and late, which correspond to the high-z and low-z contours of the reionization range established by the Plank team \citep{Planck2020a}, and intermediate z-track which corresponds to the median dependence $x_{HII}(z)$ by \cite{Glazer2018}.

Accurate computations of hydrogen and helium ionization and thermal history are crucial for solving similar tasks. We start from the early prerecombination epoch when Saha equations of recombination were applicable to all chemical components. Using the relevant basic kinetic equations, we compute the cosmological recombination using the RecFast model of an effective three-level atom. This model we apply up to $z=200$. At lower $z$ up to complete of reionization, the number density of quanta of Lyman series is not large. Therefore, hydrogen photoionization from the ground level (case A) occurs mainly (see \cite{Novosyadlyj2022}), and the kinetic equation for hydrogen ionization is simplified: 
\begin{eqnarray}
-(1+z)H\frac{dx_{HII}}{dz}=R_{HI}x_{HI}+C_i n_i x_{HI}-\alpha_{HII}x_{HII},
\label{kes}
\end{eqnarray}
where $\alpha_{HII}$ is the photorecombination rate, $R_{HI}=R_{HI}(T_{CMB})+\alpha_{fl}R_{HI}(T_{fl})$ is the photoionization rate of hydrogen, $C_i$ is collisional ionization rate by electron or/and proton, and $n_i$ is the number density of corresponding particles.   
 
Equation (\ref{kes}) for hydrogen and similar for helium we solve numerically together with the equations of expansion of the Universe and the energy balance for the baryonic component:
\begin{eqnarray}
&&\hskip-0.5cm H=H_0\sqrt{\Omega_r(1+z)^4+\Omega_m(1+z)^3+\Omega_{\Lambda}(1+z)^{3(1+\omega_{de})}},  \label{H}\\
&&\hskip-0.5cm -\frac{3}{2}n_{tot}k_B(1+z)H\frac{dT_b}{dz} = \Gamma_{C_{cmb}} +  \Gamma_{C_{fl}} +  \Gamma_{phi}+\nonumber \\
&&\hskip-0.5cm \Gamma_{phdH_2}+\Gamma_{H^-H}-\Lambda_{ad}- \Lambda_{ff}-\Lambda_{phr}-\Lambda_{21cm}-\Lambda_{H_2}- \label{Tb}\\
&&\hskip-0.5cm \Lambda_{ex}-\Lambda_{H^+H}-\Lambda_{H^-e}-\Lambda_{H^-H}-\Lambda_{ci}-\Lambda_{cdH_2}-\Lambda_{dr},\nonumber 
\end{eqnarray}
where $\Gamma_{C_{cmb}}$ is the Compton heating by CMB due to free electrons, $\Gamma_{C_{fl}}$ is the same by the first light, $\Gamma_{phi}$ is the heating by the photoionization, $\Gamma_{phdH_2}$ is the heating by the photodissociation of H$_2$ and HD, 
$\Gamma_{H^-H}$ is the heating due to reactions H$^-$ + H $\rightarrow$ H$_2$ + e,
$\Lambda_{ad}$ is the adiabatic cooling,
$\Lambda_{ff}$ is the cooling via bremsstrahlung (free-free) emission,
$\Lambda_{phr}$ is the recombination cooling,
$\Lambda_{21cm}$ is the cooling via excitation of hydrogen 21-cm line,
$\Lambda_{H_2}$ is the cooling due to collisional excitations of lines of H$_2$,
$\Lambda_{ex}$ is the cooling due to collisional excitation of HI, HeI and HeII,
$\Lambda_{H^+H}$ is the cooling by the reaction H$^+$ + H –> H$_2$ + $\gamma$, 
$\Lambda_{H^-e}$ is the cooling due to collisional deionization H$^-$ + e $\rightarrow$ H + 2e,
$\Lambda_{H^-H}$ is the cooling due to collisional deionization H$^-$ + H $\rightarrow$ 2H + e,
$\Lambda_{ci}$ is the cooling due to collisional ionization of HI, HeI and HeII,
$\Lambda_{cdH_2}$ is the cooling due to collisional dissociation of H$_2$,
$\Lambda_{dr}$ is the cooling due to dielectron recombination.
Expressions for all heating/cooling functions and their sources are presented in Appendix, equations (\ref{GCcmb})-(\ref{mfad}). 

The publicly available codes RecFast\footnote{http://www.astro.ubc.ca/people/scott/recfast.html} and  DDRIV1\footnote{http://www.netlib.org/slatec/src/ddriv1.f} have been used in the general code H21cm.f, which was designed for integrating the system of equations (\ref{kes})-(\ref{Tb}) in the expanding Universe over Cosmological Recombination, Dark Ages, Cosmic Dawn and Reionization epochs when the first light becomes important for the ionization and dissociation of atoms and molecules. The last equation (\ref{Tb}) is used at $z\le850$, at higher redshifts $T_b=T_r=T^0_{CMB}(1+z)$. 

In this section we consider the models of the first light with a single given temperature in which the ionization follows the red line in Fig. \ref{rei}. The parameters $\alpha_{fl}$, $a_{fl}$, $b_{fl}$ and $z_{fl}$ for three values of $T_*$ are presented in Tab. \ref{pmfl1}. The SEDs of the radiation (CMB+first light) for these models are shown in Fig. \ref{sedfl1} for $z=30,\;20,\,10$ and 6. One can see, the lower is $T_*$, the larger is energy density of the first light at $\nu<\nu_{L_c}$ and slower is its evolution. It is caused by increase of the spectrum steepness after the hydrogen potential of ionization (right vertical dotted line in each panel) for lower $T_{fl}$, and, accordingly, a lower number of ionizing quanta. Therefore, there is degeneration: the first light models with different SEDs and time evolutions can result in the same evolution track of fraction $x_{HII}(z)$ during reionization.

But the thermal history of the plasma can be quite different for these models of the first light. To compute it we integrate eq. (\ref{Tb}) taking into account the main heating/cooling processes listed in the Appendix. 
Their contributions to the variation of temperature ($\left(dT/dz\right)_i\equiv2\Gamma_i/\left(3n_{tot}k_B(1+z)H(z)\right)$, $\left(dT/dz\right)_i\equiv2\Lambda_i/\left(3n_{tot}k_B(1+z)H(z)\right)$) are shown in Fig. \ref{hc}. It illustrates the well-known fact: during the Dark Ages epoch the temperature of baryonic matter is determined by competition between adiabatic cooling and heating via Compton scattering of CMBR on free electrons. While during the Cosmic Dawn and Reionization epochs the heating/cooling due to photoionization by the first light and photorecombination, the Compton scattering of first light on free electrons as well as free-free transitions and inverse Compton effect becomes essential. As we see, for the same reionization history, they depend strongly on the SED of the first light. The evolution of kinetic temperature of baryonic matter from the Dark Ages up to complete hydrogen reionization is shown in Fig. \ref{figTb}. One can see, that it depends on energy density and SED of the first light as well as on their temporal variation. That will be manifested in the position and intensity of the 21-cm line from the corresponding epochs.

\section{Absorption line 21-cm from Dark Ages}

\begin{figure*}
\includegraphics[width=0.32\textwidth]{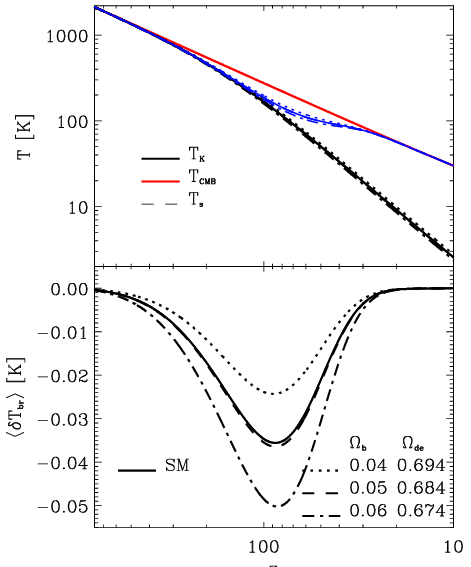}
\includegraphics[width=0.32\textwidth]{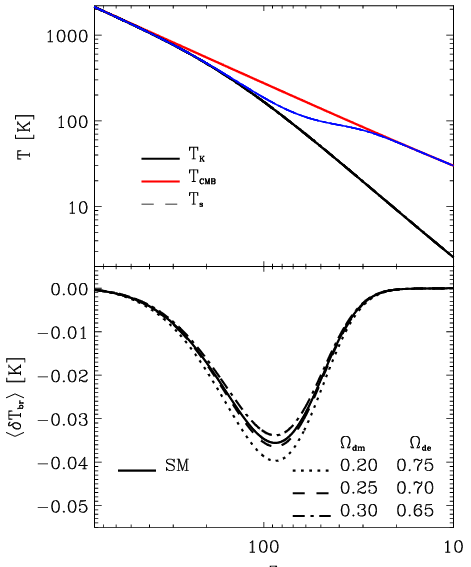}
\includegraphics[width=0.32\textwidth]{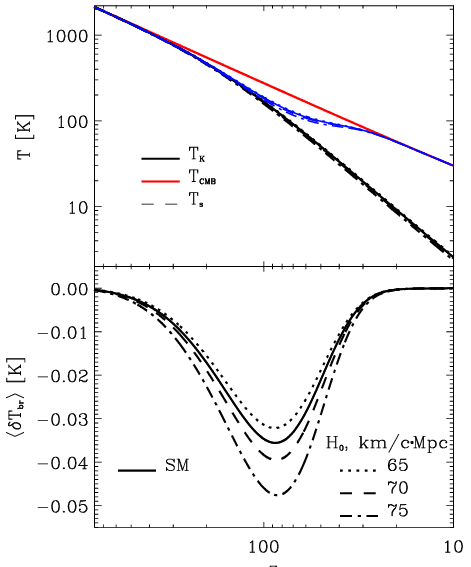}
\caption{Dependence of profile of H 21-cm line from the Dark Ages on the values of cosmological parameters: $\Omega_b$ when $\Omega_{dm}$ and $\Omega_{K}$ are unchanged (left), $\Omega_{dm}$ when $\Omega_{b}$ and $\Omega_K$ are unchanged (middle), $H_0$ when $\Omega$'s are unchanged (right). In all cases $\Omega_b+\Omega_{dm}+\Omega_{\Lambda}+\Omega_K=1$.}
\label{scp}
\end{figure*}

The signal in the redshifted line 21-cm of neutral hydrogen from the Dark Ages could be a source of information about the hydrogen state since the line's intensity depends on the number density of neutral hydrogen fraction and the kinetic temperature of baryonic matter. At $z<850$, the adiabatic cooling begins to prevail over the Compton thermalization by CMBR and the kinetic temperature decreases faster than the CMB temperature.  
It is noticeable in the logarithmic scale of Fig. \ref{figTb} at $z<400$. The spin temperature\footnote{The excitation temperature of the hyperfine transition}, which defines the populations of hydrogen hyperfine structure levels at this time and results from the kinetic equation, is as follows \citep{Field1958,Novosyadlyj2020}:  
\begin{equation}
T_s=T_b\frac{T_{cmb}+T_0}{T_b+T_0}=\frac{(1+x_c)T_{cmb}}{T_b+x_cT_{cmb}}, \quad T_0=\frac{h_P\nu_{21}C_{10}}{k_B A_{10}},
\label{Tsda}
\end{equation}
where $x_c\equiv T_0/T_{cmb}$ is called the collision coupling parameter, $\nu_{21}$ is the frequency of the 21-cm line, $A_{10}$ is the Einstein coefficient of spontaneous transition, $C_{10}$ is the collisional deactivation rate by electrons, protons and neutral hydrogen atoms, $h_P$ and $k_B$ are Planck and Boltzmann constants correspondingly. Since the hyperfine structure line frequency $\nu_{21}$ is in the Rayleigh-Jeans range of CMBR it is comfortable to use a brightness temperature $T_{br}$ instead intensity: $I_{\nu}=2k_BT_{br}\nu^2/c^2$. Since the useful signal is the difference of the redshifted intensities $\delta I_{\nu}=(I_{\nu}-I_{\nu}^{cmb})/(1+z)$ in any point of the sky the radiation transfer equation gives the expression for the differential brightness temperature in the line \citep{Madau1997,Zaldarriaga2004,Furlanetto2006,Pritchard2012} 
\begin{equation}
\delta T_{br}=\frac{T_s-T_{cmb}}{1+z}(1-e^{-\tau_{\nu_{21}}}),
\label{e_dTbr}
\end{equation}
where the optical thickness $\tau_{\nu_{21}}$ of the line forming bulk is as follows \citep{Field1959,Barkana2001,Zaldarriaga2004}
\begin{eqnarray}
\hskip-0.7cm\tau_{\nu_{21}}=\frac{3c^3h_PA_{10}n_{HI}}{32\pi k_B\nu_{21}^2T_sH(z)}=8.6\cdot10^{-3}[1+\delta_b(z,\mathbf{n})]x_{HI}\nonumber\\
\hskip0.5cm\times\left[\left(\frac{0.15}{\Omega_m}\right)\left(\frac{1+z}{10}\right)\right]^{\frac{1}{2}}
\left(\frac{\Omega_bh}{0.02}\right)\left[\frac{T_{cmb}(z)}{T_s(z)}\right]
\label{tau}
\end{eqnarray}
Here $\delta_b(z,\bf{n})$ is the density fluctuation of baryonic matter at redshift $z$ and direction in the sky $\bf{n}$.
Taking into account that the line profile at any $z$ of Dark Ages is caused by thermal processes in baryonic gas and expansion of the Universe, and small optical thickness ($\tau_{\nu_{21}}\ll1$), we can obtain the expression for sky averaged (or global) signal in the hyperfine line 21-cm of neutral hydrogen \citep{Madau1997,Zaldarriaga2004,Furlanetto2006,Pritchard2012}
\begin{eqnarray}
\hskip-0.7cm&&\delta T_{br}(z)=23x_{HI}(z)\left[\left(\frac{0.15}{\Omega_m}\right)\left(\frac{1+z}{10}\right)\right]^{\frac{1}{2}}
\left(\frac{\Omega_bh}{0.02}\right)\left[1-\frac{T_{cmb}(z)}{T_s(z)}\right]\nonumber\\
\hskip-0.7cm&&
\label{dTbr}
\end{eqnarray}
in units of mK. The multiplier $[1+\delta_b(z,\mathbf{n})]$, which is in eq. (\ref{tau}), after sky averaging became 1. The line depth at any $z$ is proportional to the difference $T_{cmb}-T_s$, which is largest at $z\approx50-100$, where the collisional processes are the most effective (Fig. \ref {scp}).
\begin{figure*}
\includegraphics[width=0.32\textwidth]{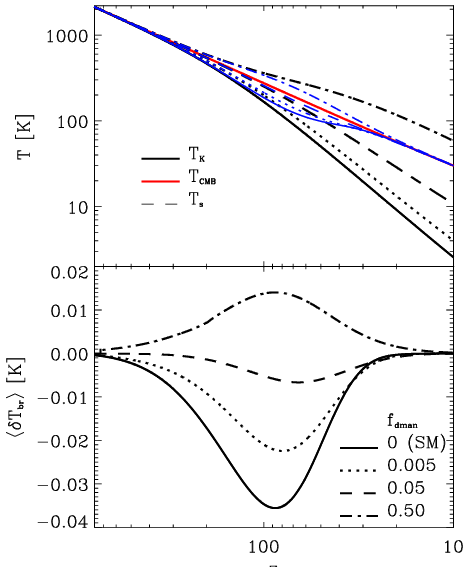}
\includegraphics[width=0.32\textwidth]{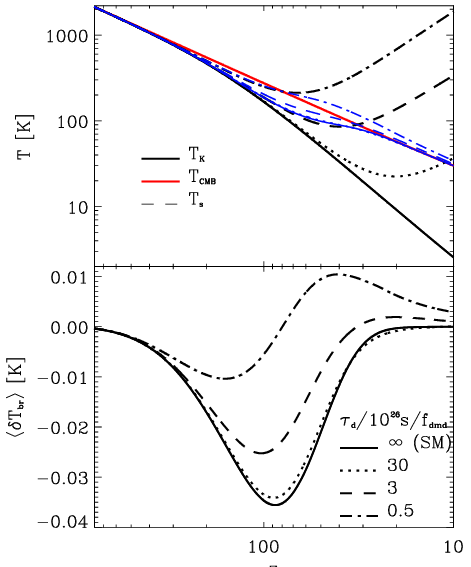}
\includegraphics[width=0.32\textwidth]{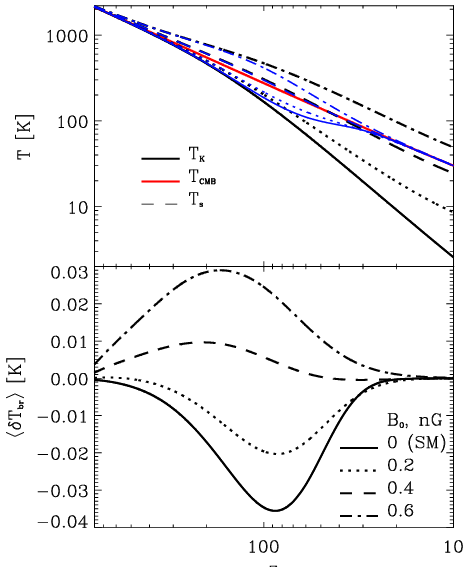}
\medskip

\includegraphics[width=0.32\textwidth]{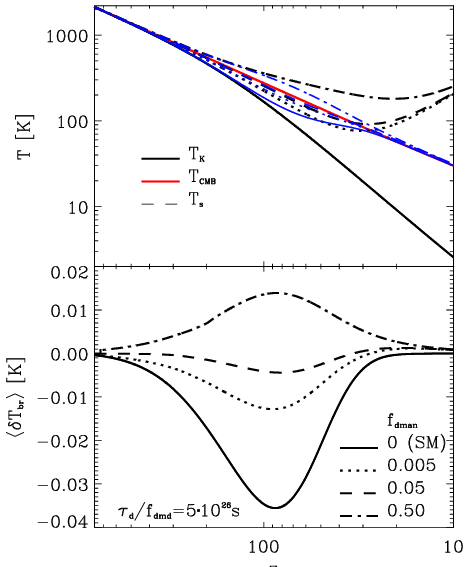}
\includegraphics[width=0.32\textwidth]{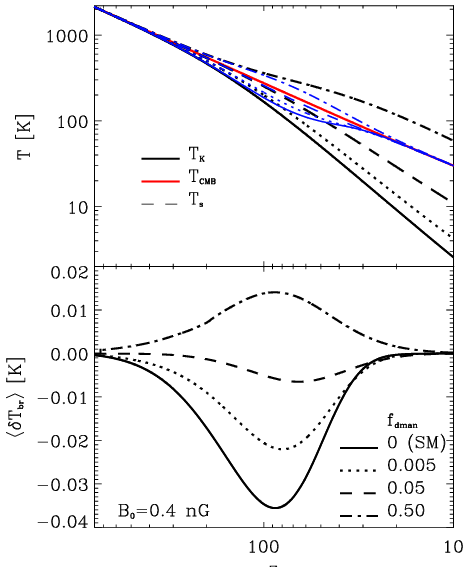}
\includegraphics[width=0.32\textwidth]{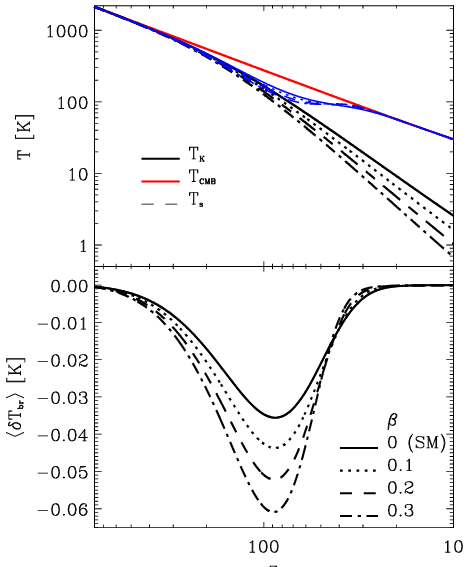}
\caption{Dependence of profile of H 21-cm line from Dark Ages on the heating and cooling in non-standard cosmology: heating by self-annihilating dark matter particles (top left), heating due to the decaying of dark matter particles (top middle), heating due to the decaying of primordial helical magnetic field (top right), heating by self-annihilating and decaying dark matter particles (bottom left), heating by self-annihilating dark matter particles and decaying of primordial magnetic field (bottom middle), cooling by interaction with cold dark matter particles (bottom right).}
\label{shc}
\end{figure*} 
The expression (\ref{dTbr}) shows an apparent dependence on some cosmological parameters, such as $\Omega_m$, $\Omega_b$, and h. Implicit dependence on these and other cosmological parameters is also in the values of $x_{HI}$, $T_s$ and the thermal history of baryonic matter. 

We compute the ionization and thermal history of baryonic matter in the Dark Ages for different cosmological parameters, and differential brightness temperature in the hyperfine line 21-cm of neutral hydrogen. The cosmological parameters were varied around ones following the final data release of the Planck Space Observatory \citep{Planck2020b}: the Hubble constant $H_0=67.36$ km/s/Mpc, the mean density of baryonic matter in the units of critical one $\Omega_b=0.0493$, the mean density of dark matter $\Omega_{dm}=0.266$, the mean density of dark energy $\Omega_{\Lambda}=0.6847$, the current temperature of cosmic microwave radiation $T^0_{CMB}=2.7255$ K. The space curvature of the fiducial model $\Omega_K=0$. We use also the primordial helium abundance $Y_p=0.2446$ \citep{Peimbert2016} and deuterium fraction $y_{Dp}=2.527\cdot10^{-5}$ \citep{Cooke2018} which well agree with the posteriors mean of \cite{Planck2020b}.  

Fig. \ref{scp} shows the sensitivity of the profile of the absorption line 21-cm formed in the Dark Ages to the variation of cosmological parameters $\Omega_b$, $\Omega_{dm}$, $\Omega_{\Lambda}$ and $\Omega_K$, which strongly satisfy the cosmological equation $\Omega_b+\Omega_{dm}+\Omega_{\Lambda}+\Omega_K=1$, as well as $H_0$. First of all, we can state that variations of $\Omega_{\Lambda}$ and $\Omega_K$ with unchanged $\Omega_b$ and $\Omega_{dm}$ do not affect the line $z$-profile at all (not shown). This result is expected in the framework of standard cosmology. While the variations of $\Omega_b$, $\Omega_{dm}$ and $H_0$ change the depth and width of $z$-profile of the line. Rising $\Omega_b$ increases the depth of the line (decreases the $\delta T_{br}^{min}$) via a larger number density of absorbers, and larger effectiveness of collisional processes in the deactivation of exited hyperfine level which stronger pull the spin temperature $T_s$  to $T_b$. And contrary, the increase of $\Omega_{dm}$ decreases the depth of the line via the Hubble expansion rate $H(z)$ in the denominator of eq. (\ref{tau}). It should be noted, that 10\% variation of $\Omega_b$ results in $\sim17\%$ variation of depth of the redshifted absorption line 21-cm, but 10\% variation of $\Omega_{dm}$ results in about 4.3\% variation of line depth. The variation of the line $z$-profile over the variation of Hubble constant $H_0$ is shown in the right panel of Fig. \ref{scp}: increasing of $H_0$ increases the depth of the line. Here 10\% variation of $H_0$ results in 28\% variation of minimal $\delta T_{br}$. In all cases, the minima are in the range $87\le z\le 93$ ($\sim15-16$ MHz). The variations of $\Omega_b$, $\Omega_{dm}$ and $H_0$ we deliberately made large for illustrative purposes. The 2$\sigma$ uncertainties of these parameters which are defined from the Planck measurements \citep{Planck2020a} are 4.5\% for $\Omega_b$, 5.2\% for $\Omega_{dm}$ and 1.6\% for $H_0$. The variations of the absorption line for them are 8.2\%, 2.5\% and 4.3\% accordingly. Figure \ref{scp} illustrates also the degeneration of dependences of $z$-profile of the redshifted absorption 21-cm line on cosmological parameters $\Omega_b$, $\Omega_{dm}$ and $H_0$. Nevertheless, the detection of the redshifted 21-cm absorption line of neutral hydrogen at frequencies 10-30 MHz from the Dark Ages could be useful to refine the values of these cosmological parameters.

The Dark Ages absorption line 21-cm is sensitive also to the possible additional mechanisms of heating/cooling which appear in the non-standard cosmological models. We set their key parameters keeping the reionization history in the $2\sigma$-range of median Planck's $x_{HII}(z)$ (fig. \ref{rei}). The results of computations of $T_b$, $T_s$ and $\delta T_{br}$ for non-standard cosmological models with annihilating and decaying dark matter and decaying of stochastic background primordial magnetic field are presented in Fig. \ref{shc} (top row). We use the models of annihilating and decaying dark matter following \cite{Chluba2010,Liu2018} accordingly. It was supposed that the mass of dark matter particles $m_{dm}=100$ GeV and the thermally averaged product of the cross-section and relative velocity of the annihilating DM particles\footnote{The rate of annihilation DM particles is proportional to the product of undefined parameters $f_{dman}\langle\sigma v\rangle/m_{dm}$ (see (33) in Appendix).} $\langle\sigma v\rangle=10^{-29}$ cm$^3$s$^{-1}$. The heating functions of baryonic matter by annihilating and decaying DM particles are presented in Appendix, formulae (\ref{dman}) and (\ref{dmd}) accordingly. The presented results for models with annihilating dark matter particles are obtained for three values of fraction of the released energy, which is deposited into the intergalactic medium $f_{dman}=$0.5\%, 5\% and 50\%. The fractions of the deposited energy that goes into the heating and ionization of atoms we computed according to prescription by \cite{Chluba2010} which gives $\approx1/3$ for each channel of deposition, heating, ionization and excitation. Such models of dark matter keep all properties of cold dark matter, so, they match well the most observational data like the fiducial model does. The profiles of the 21-cm line in the models with annihilating and decaying dark matter are noticeably different which is due to the fact that in the first case the heating function is proportional to $(1+z)^6$, and in the second one $\propto(1+z)^3$.

The additional heating mechanisms such as annihilation or decay of dark matter particles into gamma quanta, electron-positron or other charged particle-antiparticle pairs pull the kinetic temperature of baryonic gas and spin temperature to CMB one at $z<400$, which decreases the depth of the absorption line and shifting its bottom to higher redshift. At the highest values of the fraction of annihilating dark matter and the lowest values of the lifetime of decaying dark matter the absorption line turns into the emission one.  

In the top right panel of Fig. \ref{shc} we show how evolutions of $T_b$, $T_s$ and $\delta T_{br}$ depend on the r.m.s. amplitude $B_0$ of the primordial magnetic fields. We suppose they heat the baryonic matter in the post-recombination Universe due to the decaying of magnetic turbulence and the ambipolar diffusion following \cite{Sethi2005,Chluba2015,Minoda2019}. The heating functions for them (\ref{mfdt}) and (\ref{mfad}) are in Appendix. The \cite{Planck2015} data constrain the r.m.s. amplitude of the primordial magnetic fields at the nano-Gauss level, therewhy we set $B_0=$0.2, 0.4 and 0.6 nG. When $B_0$ increases the depth of the absorption line decreases and disappears when $B_0\approx0.3$ nG and turns into the emission one for larger $B_0$. The amplitude of emission 21-cm line for $B_0=$0.6 nG reaches $\approx30$ mK at $z=170$. We can also note the similarity of the profiles of the 21-cm line in models with a decaying magnetic field and annihilating/decaying dark matter with the corresponding values of the parameters of these models.

The real Universe can consist of a few sorts of dark matter particles, primordial stochastic background magnetic field and everything else that is already in the standard model. We present the evolution of $T_b$, $T_s$ and $\delta T_{br}$ in the models with annihilating ($f_{dman}=$0.5\%, 5\%, 50\%) and decaying ($\tau_{dm}/f_{dmd}=5\cdot10^{26}$ s) dark matter particles in the bottom left panel of Fig. \ref{shc}, and annihilating dark matter particles ($f_{dman}=$0.5\%, 5\%, 50\%) and decaying of primordial magnetic field with $B_0$=0.4 nG (bottom middle). 

The general trend of changing the profile of the 21-cm line of neutral hydrogen, which is formed during the Dark Ages in the models with additional heating and ionization, compared to the profile in the standard model, is a decrease in the depth of the absorption line, or even a transition to emission, determined by the specific parameters of the models. The opposite trend can be expected in models with additional cooling.

In the bottom right panel of Fig. \ref{shc} the baryon matter temperature $T_b(z)$, spin temperature $T_s(z)$ and $z$-profiles of the redshifted 21-cm line are presented for additional cooling modeled as $(1+\beta)\Lambda_{ad}$ with $\beta$=0.0, 0.1, 0.2 and 0.3, where $\Lambda_{ad}$ is the adiabatic cooling function (see Appendix). It shows that 10\% additional cooling of baryonic matter at $30\le z\le300$ results in $\sim30\%$ increase of absorption line depth. For example, the possibility of such excess cooling of the cosmic gas induced by its weak interaction with the dark matter \citep{Munoz2015} has been discussed by \cite{Barkana2018} for an explanation of the unexpected deep absorption line 21-cm detected by EDGES from Cosmic Dawn epoch \citep{Bowman2018}. Another possible reason could be the decoupling of the temperature of the baryon component from the temperature of the CMB due to the decrease in the concentration of free electrons during the Dark Ages.

\section{Hyperfine line 21-cm in Cosmic Dawn }

The populations of the hyperfine structure levels of the hydrogen ground state at $300\le z\le30$ are caused by CMB radiation and collisions of neutral hydrogen with atoms, protons and electrons. As the Universe expands, the efficiency of collisions decreases, the spin temperature approaches the CMB temperature ($T_s\rightarrow T_{cmb}$), and the 21-cm absorption line disappears at $z\sim30$, which Figs. \ref{scp}-\ref{shc} illustrate well. 

The appearance of the first extra light from the forming stars and galaxies symbolizes the beginning of the Cosmic Dawn epoch. The Universe starts to fill by $Ly\alpha$ quanta, which changes the population levels of the hyperfine structure of the ground state ($n = 1$) of atomic hydrogen through transitions between the levels of the hyperfine structure of the first excited level ($n = 2$) in such a way that $T_s$ approaches $T_b<T_{cmb}$ again, and the absorption line appears again. This effect was first predicted by the Dutch physicist Siegfried A. Wouthuysen in 1952 \citep{Wouthuysen1952} and studied in detail theoretically by George B. Field, American astrophysicist, in the late 50s \citep{Field1958,Field1959}. This phenomenon is called the Wouthuysen-Field effect or coupling and is described with different levels of detalization in a lot of reviews and textbooks, including those mentioned in the introduction. The SED of radiation of the first sources evolves increasing the ratio of the number density of Lyman continuum ($LyC$) quanta to $Ly\alpha$ ones and the appearance of X-ray radiation. It results in ionization and heating of hydrogen as shown in Fig. \ref{rei} and \ref{figTb}. The mainstream approach to describe these very complicated processes is based on the phenomenological concepts such as star formation efficiency, $Ly\alpha$ efficiency, ionizing efficiency, X-ray efficiency,the minimal virial temperature of star formation halo and a few more like these. The dependencies of the global 21-cm signal on them are analysed by \cite{Pritchard2010,Mirocha2013,Mirocha2015,Cohen2017,Monsalve2017,Monsalve2018,Monsalve2019}.

\begin{figure}
\includegraphics[width=0.49\textwidth]{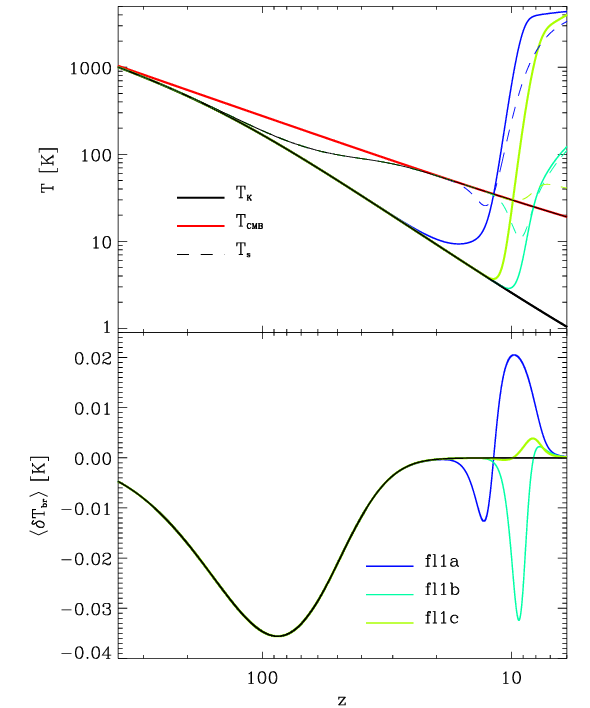}
\caption{Evolution of spin (top panel) and differential brightness (bottom panel) temperatures in the redshifted hyperfine line 21-cm of atomic hydrogen in the fl1a, fl1b and fl1c models of the first light with parameters presented in Tab. 1.}
\label{prof_flI}
\end{figure}
\begin{figure}
\includegraphics[width=0.5\textwidth]{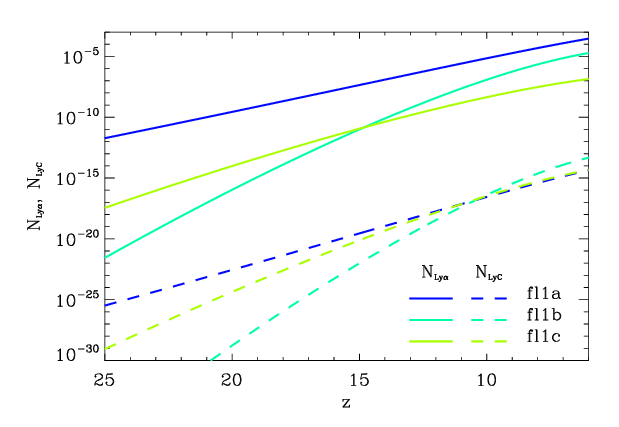}
\caption{Number densities of $N_{Ly\alpha}$ $Ly\alpha$-photons and number of ionizations of hydrogen per second $N_{LyC}$ by $LyC$-photons for Cosmic Dawn and Reionization epochs models of the first light with parameters presented in Tab. 1.}
\label{NLa1}
\end{figure}
\begin{figure*}
\includegraphics[width=0.32\textwidth]{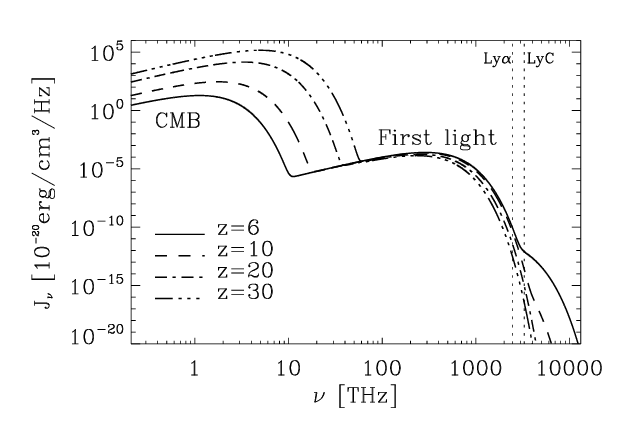}
\includegraphics[width=0.32\textwidth]{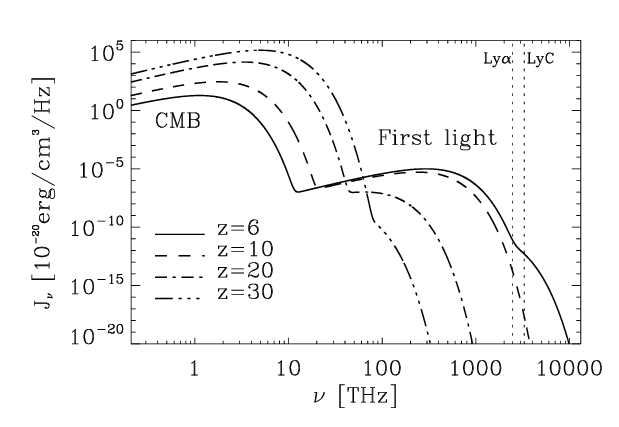}
\includegraphics[width=0.32\textwidth]{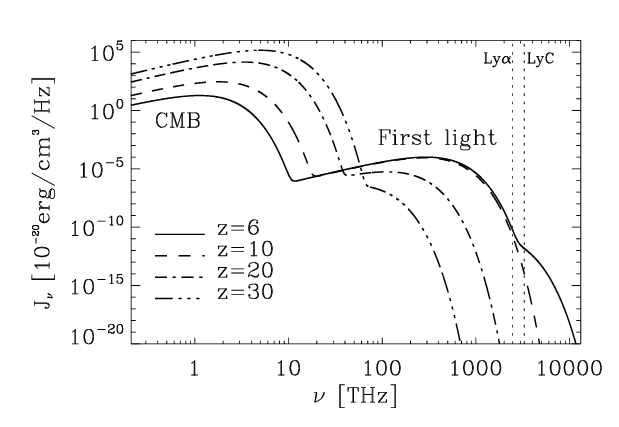}
\caption{The models of SED of radiation fl2a, fl2b and fl2c (from left to right) for Cosmic Dawn and Reionization epochs with parameters presented in Tab. 2.}
\label{sedfl2}
\end{figure*}

In this section we study the dependencies of the global 21-cm signal from the Cosmic Dawn epoch based on physical approaches and parameters: averaged energy distribution of the first light, mean energy density of $Ly\alpha$ and ionizing $LyC$ radiation and mean number density of neutral hydrogen $n_{HI}$ which match existing observations. Since the first sources of light were protostellar halos, then first stars, the first globular cluster it is natural to assume the SED of the first light as a sum of thermal distributions with some effective temperatures $T_{fl}^{(i)}$ which depend on the redshift as it was proposed in section 2, eqs. (\ref{jnu})-(\ref{tfl}). 

We compute the differential brightness temperature for the fiducial cosmological model and different models of the first light using eq. (\ref{dTbr}) with spin temperature 
\begin{equation}
T_s^{-1} = \frac{T_{cmb}^{-1} + x_c T_b^{-1} + x_\alpha T_c^{-1}}{1+x_c+x_\alpha}, \quad 
x_\alpha \equiv\frac{8\pi c^2\Delta\nu_\alpha}{9A_{10}\nu_\alpha^2}S_\alpha J_\alpha,
\label{TsCD}
\end{equation}
where $x_\alpha$ is $Ly\alpha$ coupling parameter, $\nu_\alpha$ is the frequency in the  $Ly\alpha$ line, $\Delta\nu_\alpha$ is its half-width, $S_\alpha$ is scattering function of $Ly\alpha$ quanta and $J_\alpha$ the energy density of them, $T_c$ is colour temperature in the line. $J_\alpha$ is computed using eqs. (\ref{jnu})-(\ref{tfl}),  $S_\alpha$ and $T_c$ are computed using the analytic approximation formulae (40)-(42) from \cite{Hirata2006}. They approximate the numerical results with an accuracy $\sim1\%$ in the range temperatures $T_b\ge2$ K, $T_s\ge2$ K and Gunn-Peterson optical depth $10^5\le \tau_{GP}\le10^7$, where $\tau_{GP}=2.08\cdot10^4x_{\rm HI} (1+z)^{3/2}$ for fiducial cosmological model. The last formula shows that at the redshift of complete reionization at $z\ge6$ the accuracy of the approximation is worse, but there line disappears since $x_{HI}\rightarrow0$.

Let`s suppose that SED of the first light is described by a single Planck function with effective temperature $T_{fl}(z;T_*,a_{fl},b_{fl},z_{fl})$ and dilution coefficient $\alpha_{fl}$ in (\ref{jnu})-(\ref{tfl}). The parameters $a_{fl}$, $b_{fl}$, $z_{fl}$ and $\alpha_{fl}$ are defined for given $T_*$ in such way to obtain the median value of $\overline{x}_{HII}(z)$ \citep{Glazer2018}, shown by thick red solid line in Fig. \ref{rei}. They are presented in Tab. 1  for  $T_*=$5000 K, 10000 K, 20000 K. The evolution of SED for such toy models of the first light is shown in Fig. \ref{sedfl1}. Slow evolution of energy density of the first light in the models fl1a means that the first sources in large number must appear at very large redshifts that looks unrealistic in the framework of standard cosmology. In  Fig. \ref{prof_flI} we present the evolution of $T_b$, $T_s$ (top panel), and differential brightness temperature in the redshifted 21-cm line $\delta T_{br}$ (bottom panel) for these three models of the first light. One can see how different SEDs of the first light  result in quite different profiles of $\delta T_{br}(z)$. To explain them, we present in Fig. \ref{NLa1} the evolution of the number  density of $Ly\alpha$ quanta, $N_{Ly\alpha}$, and the number of ionization per second by $LyC$ quanta, $N_{LyC}$. The absorption line starts to form (left wing) when the number density of $Ly\alpha$ quanta reaches some critical value, and starts to disappear (right wing) when $LyC$ quanta heat the baryonic matter and $T_b\rightarrow T_{cmb}$. When $T_b>T_{cmb}$ $Ly\alpha$ quanta pull $T_s$ to $T_b$, that result in the appearance of an emission line.

Now we suppose that the SED of the first light is described by two Planck functions with effective temperatures $T_{fl}^{(i)}(z;T_*^{(i)},a_{fl}^{(i)},b_{fl}^{(i)},z_{fl}^{(i)})$ and dilution coefficients $\alpha_{fl}^{(i)}$ with $i=1,\,2$ in (\ref{jnu})-(\ref{tfl}). Here we put $T_*^{(1)}=5000$ K and $T_*^{(2)}=20000$ K. We define the values of the remaining parameters to obtain the early, middle, and late reionization, as shown by two turquoise lines and a red line in Fig. \ref{rei}. They are presented in Tab. 2. 
\begin{table}
\begin{center}
\caption{Parameters of models of the first light 2.}  
\begin{tabular} {c|ccccc}
\hline
\hline
   \noalign{\smallskip}
Model&$T_*$ (K)&$\alpha_{fl}$&$z_{fl}$&$a_{fl}$&$b_{fl}$ \\
 \noalign{\smallskip} 
\hline
   \noalign{\smallskip} 
 fl2a&5000&$1.0\cdot10^{-11}$&4.6&6.0&2.5\\
 &20000&$8.0\cdot10^{-19}$&4.8&5.8&2.5\\
    \noalign{\smallskip}    
  \hline 
 fl2b &5000&$1.0\cdot10^{-10}$&5.8&6.0&2.4\\ 
 &20000&$4.5\cdot10^{-19}$&3.8&6.0&3.8\\
    \noalign{\smallskip}    
  \hline
 fl2c &5000&$1.0\cdot10^{-11}$&4.6&6.0&2.5\\ 
 &20000&$8.0\cdot10^{-19}$&4.8&2.5&5.0\\
    \noalign{\smallskip}    
  \hline  
  \hline
\end{tabular}
\end{center}
\label{pmfl2}
\end{table}
The SED of radiation (\ref{jnu}) for these models of the first light at Cosmic Dawn and Reionization epochs are shown in Fig. \ref{sedfl2}. As in the previous case, the very slow evolution of energy density of the first light in the models fl2a means that the first sources in large number must appear at very large redshifts that looks unrealistic in the framework of standard cosmology. 
\begin{figure}
\includegraphics[width=0.49\textwidth]{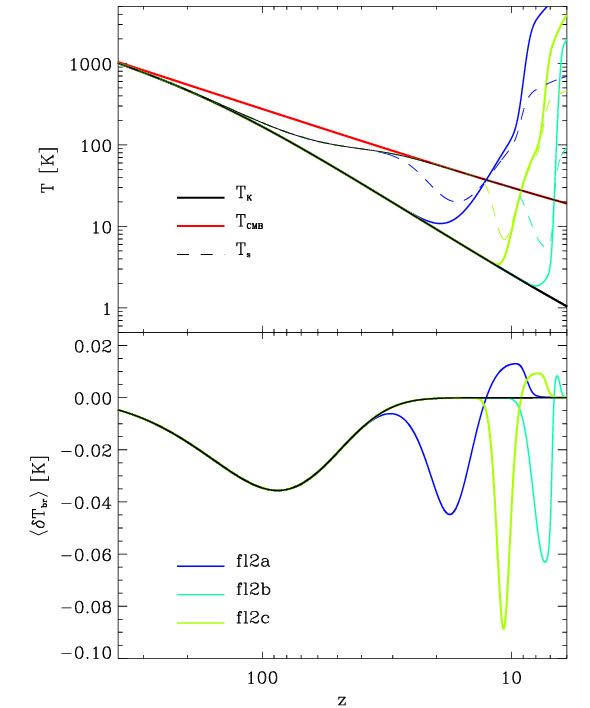}
\caption{Evolution of spin (top panel) and differential brightness (bottom panel) temperatures in the redshifted hyperfine line 21-cm of atomic hydrogen in the fl2a, fl2b and fl2c models of the first light with parameters presented in Tab. 2.}
\label{prof_fl2}
\end{figure}
\begin{figure}
\includegraphics[width=0.5\textwidth]{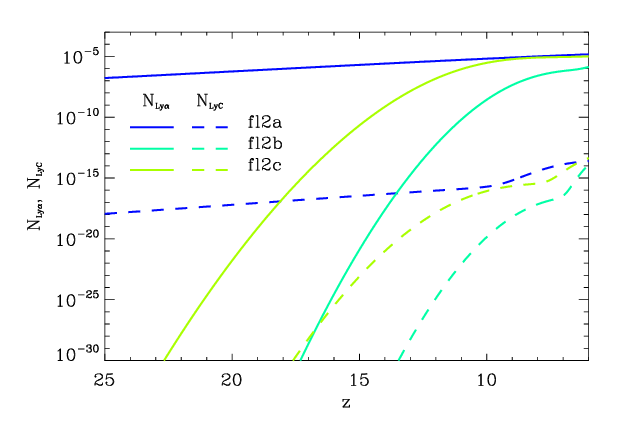}
\caption{Number densities of $N_{Ly\alpha}$ $Ly\alpha$ photons and number of ionizations of hydrogen per second $N_{LyC}$ by $LyC$-photons for Cosmic Dawn and Reionization epochs with parameters presented in Tab. 2.}
\label{NLa2}
\end{figure}
Evolutions of spin and differential brightness temperatures in the redshifted hyperfine line 21-cm of atomic hydrogen in the fl2a, fl2b and fl2c models of the first light with parameters presented in Tab. 2 are shown in Fig. \ref{prof_fl2}. We can see that absorption lines in these models appear at $z=$17.7 (at the central frequency 76 MHz) with $\delta T_{br}^{min}=$-0.045 mK, $z=$10.8 (120 MHz) with $\delta T_{br}^{min}=$-0.089 mK and $z=$7.3 (170 MHz) with $\delta T_{br}^{min}=$-0.063 mK accordingly. Much weaker emission lines are at $z=$9.7 ($\delta T_{br}^{max}=$0.013 mK), $z=$7.8 with ($\delta T_{br}^{max}=$0.0093 mK) and at $z=$6.6 ($\delta T_{br}^{max}=$0.0083 mK). The number densities $N_{Ly\alpha}$ of $Ly\alpha$-photons and the number of ionizations of hydrogen per second $N_{LyC}$ by $LyC$-photons are presented in Fig. \ref{NLa2} for these models of the first light. Together with the evolution of $T_b$ in the top panel of Fig. \ref{prof_fl2} they explain the appearance of the absorption and emission lines caused by Wouthuysen-Field coupling.

It should be noted that line profiles similar to those obtained here can be found in   \cite{Pritchard2010,Mirocha2013,Mirocha2015,Cohen2017,Monsalve2017,Monsalve2018,Monsalve2019} for other parameterizations of the first light models.

The model of the first light (\ref{tfl}) used here has a sufficient degree of freedom to simulate its different history and the different $N_{Ly\alpha}/N_{LyC}$ ratio, however, we were unable to obtain a global signal in the 21 cm line larger than 100 mK in the framework of standard cosmology. This can be explained by reference to the observed limitations on reionization and the limitation of the number of models with other values of the parameter $T_*$. Only with additional cooling we can obtain signal $\sim200$ mK, which, however, is still far from the announced result $\sim530$ mK obtained in the EDGES experiment \citep{Bowman2018} at the centre frequency 78 MHz, that corresponds $z\approx17$. The results of the SARAS3 experiment, which were published last year \citep{Singh2022}, however, rejected the EDGES best-fit profile at 2$\sigma$ confidence level. The arbitral experiments are desirable. Some of them are being implemented now, while others are planned. 

\section{Conclusion}

We analysed the spectral features of radiation in the range 5-200 MHz formed by the hyperfine structure of the ground level  of hydrogen atoms during the Dark Ages in the range of redshift $30<z<300$, and by the ground and first excited levels during the Cosmic Dawn and Reionization epochs in the range of redshift $6<z<30$. In the $\Lambda$CDM model with post-Planck parameters the first such feature is the broad absorption line at the frequency of 16 MHz ($z=87$) with a depth $\delta T_{br}=38.6$ mK and a full width at half maximum (FWHM) of 23 MHz. The depth and FWHM show noticeable dependence on cosmological parameters, such as baryon and dark matter density parameters, $\Omega_b$ and $\Omega_{dm}$, and Hubble constant $H_0$. So, its detection can be used to refine the values of these cosmological parameters. The profile of this line is also sensitive to additional mechanisms of cooling and heating of baryonic gas in non-standard cosmology and can be used for its testing.

The second crucial spectral feature is the absorption line at frequency range $68<\nu<180$ MHz ($7<z<20$) with depth from 0 up to $\sim80$ mK and FWHM$\sim$15-40 MHz, which is caused by Wouthuysen-Field coupling between the scattering of $Ly\alpha$ radiation of the first sources on neutral hydrogen atoms and population of the hyperfine their levels. The position, depth, and FWHM drastically depend on the SED of the first light and its time history.  So, detecting this line is very important for studying the first light sources.

The third spectral feature, which appears due to the Wouthuysen-Field effect and heating of the baryonic matter by UV-radiation during the Reionization epoch, is the emission line 21 cm redshifted to the meter-size wavelengths. This emission frequency is in the range $130<\nu<190$ MHz, and the amplitude is below 20 mK. Its disappearing is caused by the complete reionization of hydrogen atoms. The detection of an emission line 21-cm redshifted to 1.5 -- 2.5 m can provide important information about the progress of hydrogen reionization in the final stage.

The dependences of the amplitudes and FWHM of these lines on the parameters of the cosmological and first-light models are strongly degenerate. Detection of all three features by tomography of the Dark Ages, Cosmic Dawn, and Reionization epochs, in the 21 cm line, would significantly reduce it. This task is a challenge even for modern advanced telescopes, receivers, and technologies for extracting a useful signal from the prevailing noise background.

We have modeled the evolution of SED of the first light by thermal sources with different effective temperatures and dilution coefficients. Since most of the cosmological scenarios of galaxies formation admit succession Pop III - Pop II - globular cluster - dwarf galaxies or like that, such an assumption makes sense. The results and conclusions obtained in such models of the first light are worthy of attention: i) the flatter the radiation energy distribution in the Lyman-continuum, the lower are the radiation energy density, the temperature of the baryon gas, and the concentration of Lyman-alpha quanta, therefore, the lower is  depth of absorption line in the frequency range of 45-200 MHz, or even its absence (fig. \ref{prof_flI}); ii) the amplitude of the absorption line in this frequency range does not exceed 100 mK (fig. \ref{prof_fl2}) if the observational limits of the reionization z-region obtained in Planck experiment \citep{Planck2020a} is taken into account; iii) the emission redshifted 21-cm line with an amplitude below 20 mK in the Reionization epoch is possible when $0.2<x_{HII}<0.99$.  

\section*{Acknowledgements}
This work was supported by the International Center of Future Science and College of Physics of Jilin University (P.R.China), and the project of Ministry of Education and Science of Ukraine ``Modeling the luminosity of elements of the large-scale structure of the early universe and the remnants of galactic supernovae and the observation of variable stars'' (state registration number 0122U001834). 

\section*{Data availability}
The data underlying this article will be shared on reasonable request to the corresponding author Bohdan Novosyadlyj (bnovos@gmail.com).

\begin{onecolumn}
\section*{Appendix A. Heating/cooling functions}

{\normalsize
 
Heating due to Compton scattering of thermal radiation with temperature $T_r$ on free electrons (\mbox{\footnotesize{Seager et al. (1999);\,Weyman\,(1965)}}):
\begin{eqnarray}
 \Gamma_C & =& \frac{4\sigma_Tk_Ba_{r}n_e}{m_ec}T^4_r\left(T_r-T_b\right) \quad {\rm erg/cm^3s}.
\label{GCcmb}
\end{eqnarray}
Adiabatic cooling (\mbox{\footnotesize{Seager et al. (1999);\,Peebles (1971, 1993)}}):
\begin{eqnarray}
\Lambda_{ad}= -3n_{tot}k_BT_bH(z)\left(1+\frac{1}{3}\frac{d\ln{(1+\delta)}}{dz}\right)\quad {\rm erg/cm^3s}. 
\label{Lad}
\end{eqnarray} 
Bremsstraglung (free-free) emission (\mbox{\footnotesize{Shapiro \& Kang (1987); Spitzer (1978)}}):
\begin{eqnarray}
\Lambda_{ff}&=& 1.426\cdot10^{-27}\sqrt{T}\Sigma_i[0.79464+0.1243\log{T/Z_i}]Z_i^2n_en_i\quad {\rm erg/cm^3s}. 
\label{Lff}
\end{eqnarray}
Heating by photoionizations (\mbox{\footnotesize{Anninos et al. (1997);\,Osterbrock (1974)}}):
\begin{eqnarray}
&&\Gamma_{phi}=4\pi n_{i}\int_{\nu_i}^{10\nu_i}\sigma_i(\nu) B_{\nu}\frac{\nu_i-\nu}{\nu}d\nu \quad {\rm erg/cm^3s},\quad \sigma_{HeII}(\nu)=1.58\cdot10^{-18}(\alpha_{HeII}^2+1)^{-4}e^{4-\frac{4}{\alpha_{HeII}\tan(\alpha_{HeII})}}/(1-e^{-\frac{2\pi}{\alpha_{HeII}}})\quad {\rm cm^2},\\
&&\sigma_{HI}(\nu)=6.3\cdot10^{-18}(\alpha_{HI}^2+1)^{-4}e^{4-\frac{4}{\alpha_{HI}\tan(\alpha_{HI})}}/(1-e^{-\frac{2\pi}{\alpha_{HI}}}),\quad
\sigma_{HeI}(\nu)=7.42\cdot10^{-18}\left[1.66(\alpha_{HeI}^2+1)^{-2.05}-0.66(\alpha_{HeI}^2+1)^{3.05}\right]\quad {\rm cm^2}.\nonumber
\label{Gphi}
\end{eqnarray}
Recombination cooling (\mbox{\footnotesize{Anninos\,et\,al.\,(1997);\,Black\,(1981);\,Spitzer\,(1978)}}):
\begin{eqnarray}
\Lambda^{(HII)}_{phr}&=& 8.7\cdot10^{-27}\sqrt{T_b}\left(T_b/10^3\right)^{-0.2}\left[1+\left(T_b/10^6\right)^{0.7}\right]^{-1}n_e n_{HII}\quad {\rm erg/cm^3s}, \\
\Lambda^{(HeII)}_{phr}&=& \left(1.55\cdot10^{-26}T_b^{0.3647}+1.24\cdot10^{-13}T_b^{-1.5}\left[1+0.3e^{-94000/T_b}\right]e^{-470000/T_b}\right)n_e n_{HeII}\quad {\rm erg/cm^3s},\\
\Lambda^{(HeIII)}_{phr}&=& 3.48\cdot10^{-28}\sqrt{T_b}\left(T_b/10^3\right)^{-0.2}\left[1+\left(T_b/10^6\right)^{0.7}\right]^{-1}n_e n_{HeIII}\quad {\rm erg/cm^3s}.
\label{Lphr}
\end{eqnarray}
Cooling via excitation of hydrogen 21-cm line (\mbox{\footnotesize{Seager et al. (1999)}}):
\begin{eqnarray}
 \Lambda_{21cm}&=&h_P\nu_{10}*\left(n_{HI_0}C_{01}-n_{HI_1}C_{10}\right) \quad {\rm erg/cm^3s}. 
\label{L21}
 \end{eqnarray}
Heating in reactions H$^-$+H$\rightarrow$H$_2$+e (\mbox{\footnotesize{Shapiro \& Kang (1987); Hollenbach \& McKee (1979)}}):
\begin{eqnarray}
\Gamma_{H^-H}&=& 1.3\cdot10^{-9}n_{H^-}n_{HI}\frac{3.53n_{HI}}{n_{HI}+n_{cr}} \quad {\rm erg/cm^3s}, \quad
 n_{cr}  = \frac{10^6/\sqrt{T}}{1.6x_{HI}\exp{\left[-(400/T)^2\right]}+1.4x_{H_2}\exp{\left[-12000/(T+1200)\right]}}\,\,\mbox{\rm cm$^{-3}$}.
\label{GH-H}
 \end{eqnarray}
Cooling due to collisional  excitation of lines of H$_2$ (\mbox{\footnotesize{Seager et al. (1999)}}):
\begin{eqnarray}
 \Lambda_{H_2}=10^{a_0+a_1x+a_2x^2+a_3x^3+a_4x^4+a_5x^5},\quad x\equiv\lg{(T/1000)}\quad {\rm erg/cm^3s}.
 \label{LH2}
\end{eqnarray}
Heating by photodissociation H$_2$ and HD (\mbox{\footnotesize{Shapiro \& Kang (1987); Coppola et al. (2011)}}):
\begin{eqnarray}
&&\Gamma_{phdH_2}= 6.41\cdot10^{-13}(n_{H2}+n_{HD})(k_{30}+k_{31})\quad {\rm erg/cm^3s},\quad k_{30}  = 6.46\cdot10^{8}\left[\alpha_{fl1}e^{-165530/T_{fl1}}+\alpha_{fl2}e^{-165530/T_{fl2}}\right], \\
&&k_{31}  = 1.27\cdot10^{8}\left[\alpha_{fl1}T_{fl1}^{0.084}e^{-159600/T_{fl1}}+\alpha_{fl2}T_{fl2}^{0.084}e^{-159600/T_{fl2}}\right].
\label{GphdH}
\end{eqnarray}
Cooling by reactions H$^+$ + H --> H$_2$ + $\gamma$ (\mbox{\footnotesize{Shapiro \& Kang (1987)}}):
\begin{eqnarray}
\Lambda_{pHI}&=&3.83\cdot10^{-39}T_b^{2.8}n_{HI}n_{HII}\quad {\rm erg/cm^3s} \quad \mbox{\rm for}\quad T_b\le 6700\, \mbox{\rm K},\quad \\
\Lambda_{pHI}&=&1.2\cdot10^{-31}T_b(T_b/56200)^{-0.6657\lg{(T_b/56200)}}n_{HI}n_{HII}\quad {\rm erg/cm^3s} \quad \mbox{\rm for}\quad T_b>6700\, \mbox{\rm K}
\label{LpHI}
\end{eqnarray}
Cooling due to collisional excitation of HI, HeI and HeII (\mbox{\footnotesize{Anninos et al. (1997), Black (1981); Cen (1992)}}):
\begin{eqnarray}
 \Lambda_{c-ex}&=&7.5\cdot10^{-19}\left(1+\sqrt{T_b/10^5}\right)^{-1}e^{-118348/T_b}n_en_{HI}
 +9.1\cdot10^{-27}\left(1+\sqrt{T_b/10^5}\right)^{-1}T_b^{-0.1687}e^{-13179/T_b}n_e^2n_{HeI}\\
 &+&5.54\cdot10^{-17}\left(1+\sqrt{T_b/10^5}\right)^{-1}T_b^{-0.397}e^{-473638/T_b}n_en_{HeII} \quad {\rm erg/cm^3s}.
\label{Lcex}
\end{eqnarray}
Cooling due to collisional deionization H$^-$ + e --> H + 2e (\mbox{\footnotesize{Shapiro \& Kang (1987)}}):
\begin{eqnarray}
\Lambda_{H^-e}&=&4.801\cdot10^{-24}T_be^{-8750/T_b}n_en_{H^-}  \quad {\rm erg/cm^3s}.
\label{LH-e}
\end{eqnarray}
Cooling due to collisional deionization H$^-$ + H --> 2H + e (\mbox{\footnotesize{Shapiro \& Kang (1987)}}):
\begin{eqnarray}
\Lambda_{H^-H}&=&6.368\cdot10^{-32}T_b^{2.17}e^{-8750/T_b}n_{HI}n_{H^-}  \quad {\rm erg/cm^3s}.
\label{LH-H}
\end{eqnarray}
Cooling due to collisional ionization of HI, HeI and HeII (\mbox{\footnotesize{Anninos et al. (1997); Shapiro \& Kang (1987); Cen (1992)}}):
\begin{eqnarray}
\Lambda_{ci}&=&2.77\cdot10^{-32}\sqrt{T_b}\left(1+\sqrt{T_b/10^5}\right)^{-1}e^{-157809.1/T_b}n_en_{HI}
+3.7\cdot10^{-32}\sqrt{T_b}\left(1+\sqrt{T_b/10^5}\right)^{-1}e^{-285335.4/T_b}n_en_{HeI}\\
&+&4.32\cdot10^{-32}\sqrt{T_b}\left(1+\sqrt{T_b/10^5}\right)^{-1}e^{-631515/T_b}n_en_{HeII}
+5.01\cdot10^{-27}T_b^{-0.1687}\left(1+\sqrt{T_b/10^5}\right)^{-1}e^{-55338/T_b}n_e^2n_{HeII}  \quad {\rm erg/cm^3s}.\nonumber
\label{Lci}
\end{eqnarray}
Cooling due to collisional dissociation of H$_2$ (\mbox{\footnotesize{Shapiro \& Kang (1987); Cen (1992)}}):
\begin{eqnarray}
 &&\Lambda_{cdH_2}=3.14\cdot10^{-21}n_{H_2}\left[e^{-102000/T_b}n_e+2.74\left(10.7e^{17950/T_b}\right)^{1+n_{HI}/n_{cr}^{(HIH_2)}}n_{HI}
 +3.0\left(11e^{16200/T_b}\right)^{1+n_{HI}/n_{cr}^{(H_2H_2)}}n_{H_2}\right],\\
 &&n_{cr}^{(HIH_2)}=10^{4.d0-0.416\lg{(T/10^4)}-0.327(\lg{(T/1.d4)})^2}, \quad
  n_{cr}^{(H_2H_2)}=10^{4.845d0-1.3\lg{(T/10^4)}+1.62(\lg{(T/1.d4)})^2}  \quad {\rm erg/cm^3s}.
\label{LcdH2}
\end{eqnarray}
Cooling due to dielectron recombination (\mbox{\footnotesize{Shapiro \& Kang (1987)}}):
\begin{eqnarray}
\Lambda_{dr}&=&1.24\cdot10^{-13}T^{-1.5}e^{-470000/T_b}\left[1+0.3e^{-94000/T_b}\right]  \quad {\rm erg/cm^3s}.
\label{Ldr}
\end{eqnarray}

Heating due to dark matter annihilation (\mbox{\footnotesize{Chluba (2010)}}):
\begin{eqnarray}
\Gamma_{dman} =2.4\cdot10^{-36}f_{dman}g_h n_H(1+z)^3\left[\frac{m_{dm}}{100\mathrm{GeV}}\right]^{-1}\left[\frac{\Omega_{dm}h^2}{0.13}\right]^2\left[\frac{\langle\sigma v\rangle}{3\cdot10^{-26}\mathrm{cm^2s^{-1}}}\right]   \quad {\rm erg/cm^3s},
\label{dman}
\end{eqnarray}
where $x_{HII} = n_{HII}/n_{H}$, $x_{HeII} = n_{HeII}/n_{He}$, $f_{He}=n_{He}/n_H$, $m_{dm}$ is mass of DM particle, $\langle\sigma v\rangle$ is the thermally averaged product of the cross-section and relative velocity of the annihilating DM particles, $f_{dman}$ is fraction of the released energy which is deposited into the intergalactic medium, $g_h=(1 + 2x_{HII} + f_{He}(1 + 2x_{HeII}))/3(1+f_{He})$ is fraction of deposited energy which going into the heating of gas. 

Heating due to decay of dark matter  (\mbox{\footnotesize{Liu \& Slatyer (2018)}}):
\begin{eqnarray}
\Gamma_{dmd}=2.2\cdot10^{-9}f_{dmd}g_h\frac{(1+z)^3}{\tau_{dm}}\left[\frac{\Omega_{dm}h^2}{0.13}\right]^2 \quad {\rm erg/cm^3s},
\label{dmd}
\end{eqnarray}
where $f_{dmd}$ is fraction of the released energy which is deposited into the intergalactic medium, $\tau_{dmd}$ is decay lifetime, $g_h$ is fraction of deposited energy which going into the heating of gas taken from \cite{Chluba2010}.

Heating due to decaying turbulence of the primordial magnetic field (\mbox{\footnotesize{Sethi \& Subramanian (2005), Chluba et al. (2015)}}):
\begin{eqnarray}
\Gamma_{mfdt} &=& 1.5\rho_{mf}H(z)[f_D(z)]^{n_B+3}\frac{ma^m}{(a+1.5\ln((1+z_{cr})/(1+z)))^{m+1}} \quad \mbox{for} \quad z<z_{cr}, \nonumber \\ 
\Gamma_{mfdt} &=& 1.5\rho_{mf}H(z)\frac{m}{a}[f_D(z)]^{n_B+3}\exp\left\{-\frac{(z-z_{cr})^2}{5000}\right\}\left(\frac{1+z_{cr}}{1+z}\right)^4\quad \mbox{for} \quad z\ge z_{cr},
\label{mfdt}
\end{eqnarray}
where $z_{cr}=1088$, $\rho_{mf} =3.98\cdot10^{-21}\left(B_0/{\text{nG}}\right)^2(1+z)^4$~J$\cdot$m$^{-3}$, $n_B=-2.9$, $a = \ln(1+t_\mathrm{d}/t_\mathrm{rec})$, $m \equiv 2(n_B+3)/(n_B+5)$, $t_\mathrm{d}/t_\mathrm{rec}=14.8/(B_0k_D)$, $k_D=(2.89\cdot10^4 h)^{1/(n_B+5)}B_\lambda^{-2/(n_B+5)}k_\lambda^{(n_B+3)/(n_B+5)}$ Mpc$^{-1}$, $\lambda = 1$ Mpc, $B_\lambda = B_{1\,\mathrm{Mpc}} =B_0$, $k_\lambda = k_{1\,\mathrm{Mpc}} = 2\pi$ Mpc$^{-1}$. The factor $f_D(z)^{n_B+3}$ describes the energy loss by the primordial magnetic field (see \cite{Minoda2019}), which we approximate as  
$[f_D(z)]^{n_B+3} \simeq 0.6897525+0.2944149\cdot10^{-3}z-0.3805730\cdot10^{-6}z^2+0.2259742\cdot10^{-9}z^3+0.6354026\cdot10^{-13}z^4$ for $z < 1178$, fixed values of $n_B$ and $k_D$  (for $z\ge1178$ $ f^{n_B+3}(z) \equiv 1$).

Heating due to ambipolar diffusion caused by the primordial magnetic field (\mbox{\footnotesize{Chluba et al. (2015), Minoda et al. (2019)}}):
\begin{eqnarray}
\Gamma_{mfad} = \frac{1-x_{HII}}{g(T_b)x_{HII}}[f_D(z)]^{2n_B+8}\left[\frac{(1+z)k_D}{3.086\cdot10^{22}}\frac{\rho_{mf}}{\rho_b}\right]^2f_L, 
\label{mfad}
\end{eqnarray}
where $x_{HII} = n_{HII}/n_{H}$, $f_L=0.8313(n_B+3)^{1.105}(1.0-0.0102(n_B+3))$, $g(T_b) = 1.95\cdot10^{11} T_b^{0.375}$~m$^3$/s/kg, $\rho_b = \rho_{cr}^{(0)}\Omega_b(1+z)^3$, $k_D=286.91(B_0/\text{nG})^{-1}$~Mpc$^{-1}$. 
} 
\end{onecolumn}

\end{document}